\documentclass[sort&compress]{elsarticle}

\usepackage{graphicx}
\usepackage{amssymb}
\usepackage{amsmath}
\usepackage{epsfig}
\usepackage{epstopdf}
\usepackage{subfig}
\journal{EPJP}
\begin{document}
\begin{frontmatter}

\title{Theoretical Exploration on the Magnetic Properties of Ferromagnetic Metallic Glass: An Ising Model on Random Recursive Lattice}

\author[sjtu]{Ran ˜Huang\corref{cor1}}
\ead{ranhuang@sjtu.edu.cn}
\author[tju]{Ling ˜Zhang}
%\ead{1334099@tongji.edu.cn}
\author[company]{Chong ˜Chen}
%\ead{insar.cchen@gmail.com}
\author[wu]{Chengjie ˜Wu}
%\ead{wuchjie@gmail.com}
\author[cas]{Linyin ˜Yan\corref{cor2}}
\ead{yanlinyin@iccas.ac.cn}
\cortext[cor1]{Corresponding author}
\cortext[cor2]{Co-corresponding author}
\address[sjtu]{School of Chemistry and Chemical Engineering, Shanghai Jiao Tong University, Shanghai 200240, China}
\address[tju]{Department of Physics, Tongji University, Shanghai 200092, China}
\address[company]{Ningbo Hongzheng Engineering Consulting Co. Ltd, Ningbo, Zhejiang 315000, China}
\address[wu]{Department of Computer Science and Engineering, Washington University in St. Louis, St. Louis, MO 63130, United States}
\address[cas]{Institute of Process Engineering, Chinese Academy of Science, Beijing 100190, China}

\begin{abstract}
The ferromagnetic Ising spins are modeled on a recursive lattice constructed from random-angled rhombus units with stochastic configurations, to study the magnetic properties of the bulk Fe-based metallic glass. The integration of spins on the structural glass model well represents the magnetic moments in the glassy metal. The model is exactly solved by the recursive calculation technique. The magnetization of the amorphous Ising spins, i.e. the glassy metallic magnet is investigated by our modeling and calculation on a theoretical base. The results show that the glassy metallic magnets has a lower Curie temperature, weaker magnetization, and higher entropy comparing to the regular ferromagnet in crystal form. These findings can be understood with the randomness of the amorphous system, and agrees well with others' experimental observations.

\end{abstract}


\end{frontmatter}

\section{Introduction}

Since the first report of Fe-based bulk metallic glass (BMG) in 1995 \cite{first}, Fe-based BMG with high glass forming ability (GFA) has drawn an intensive research interest in the last two decades. The nature of glassy state provides advantages of high fracture strength, corrosion resistance, and excellent soft-magnetic properties \cite{review,intro1,intro2,intro3}. More specifically, differentiated from the conventional ferromagnet in the crystal form, the feature of easy magnetization and demagnetization of the Fe-based BMG enables various significant applications in either scientific research or industrial engineering \cite{application}. 

Consequently, in almost every research on BMGs, the study of magnetic properties of such materials is an indispensable concern. Some phenomenological principles can be indicated with a general review of experimental works, for example, the magnetic moments per transition metal atom, the Curie temperature, and the magnetization of metallic glass are lower than in the corresponding crystal alloy \cite{1,prb,SLAWSKA-WANIEWSKA,general1,general2,AfliedCo}; The smaller GFA corresponds to the higher transition temperatures, especially the melting transition $T_{\text{M}}$, i.e. a more thermally stable system \cite{AfliedCo,xiamenU,xinjiang}; And external magnetic field can enhance the soft magnetic properties of Fe-based bulk amorphous materials \cite{shenyang}. These observations accord well to our intuitive knowledge of the amorphousness. Nevertheless, investigations on the featured magnetism of BMG have rarely been conducted on a theoretical base, especially from the aspect of thermodynamics with the fundamental concepts of the glass and crystal.

On the other hand, along with the experimental work, the simulation and computation on the metallic glass have also became an activate field \cite{MD1,MD2}. However, by this time the theoretical analysis mainly focus on the glass formation and simple thermodynamics, simulation work involves the concept of spin to study the magnetic property is rare, Ref. \cite{MD3} as an example. 

According to the above concern, in this work we conduct a theoretical exploration on the magnetic properties of ferro-BMG with an abstractive thermodynamic model of spins' crystal and glass. The classical Ising spins are modeled on a random-angled recursive lattice to present the magnetic moments arranged in an amorphous configuration. The integration of Ising spins and the structural glass model well represents the case of ferromagnetic metallic glass, while the recursive lattice technique enables exact thermodynamic calculation of the model. In this way, the spontaneous magnetization transition with Curie temperature $T_{C}$, the effects of randomness parameter, and the entropy behavior are quantitatively investigated.

\section{The Model}
Ising model was originally designed to describe the magnetization, and has been developed to study the magnetic materials for many years \cite{Ising,wiki}. The model is usually applied on lattices for numerical treatment. It should be addressed that, the model in this paper is not the \textit{spin glass} \cite{Debenedetti}: the orientation of spins only relates to the degree of magnetization, and the amorphousness of glass is represented by the configurational randomness, i.e. the system is a \textit{structural glass} with magnetic spins as particles.

\subsection{Lattice Construction}
\begin{figure}
    \centering{
	\includegraphics[width=0.48\textwidth]{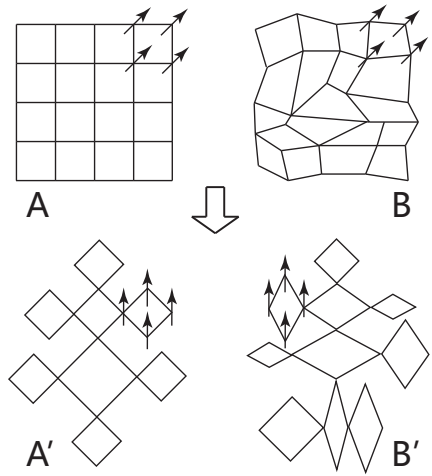}
	\caption{The recursive lattices constructed from regular lattices, fixed (A \& A') and random (B \& B') conformational respectively. The fixed recursive one is the classical Husimi lattice (A'). The lattice B' used in this paper employs rhombus units with variable angles to mimic the random lattice B. Note that in A' and B' all the edges are in the same length, however it cannot be drawn on a page in that way. Four spins are presented in each lattice to demonstrate the uniform magnetic moment orientation regardless of the structural ordering.}
	\label{fig1}
    }
\end{figure}

Among numerous methods developed to solve the Ising model, the recursive lattice has been proved to be a reliable method with the advantages of simple formulation and exact calculation \cite{Wu,exact,pdg_prl_reliable,PDG5,EJ}. Recursive lattices is constructed by connecting the units from regular lattice on the vertex in a recursive fashion, to approximate the regular lattice with the same coordination number. Such a lattice assembled by square units is called \textit{Husimi lattice} \cite{Husimi1}. Fig.\ref{fig1} illustrates how a Husimi lattice (A') is drawn from a regular square lattice (A).

For the structural glass (Fig.\ref{fig1}B), a calculable regular lattice is impossible. However, the unit cell in a recursive lattice only interacts to others on the joint vertex and its conformation is independent, this makes a individually random unit with variable angles possible. Herein, we construct a Husimi rhombus lattice with random angles to demonstrate the conformational stochasticity of structural glass (Fig.\ref{fig1}B'). In this random lattice, the conformation of each rhombus is randomized and fixed by the program, therefore locally it has a quenched randomness. 

By setting the square to be the lowest energy conformation, any randomized off-$90^\circ$ rhombus will have a higher energy, which implies a more disordered conformation deviated from the crystal. Depending on the setup of parameter, we can artificially control how ``far away" the structure is from the ``pure crystal" form. Thereafter, it is better to define our amorphous model as ``semi-glass", since it is a metastable state other than the crystal, but not reaching the upper limit of amorphousness, i.e. the ``pure glass".

\subsection{Ising model and calculation}
The Ising spins are applied on the lattices as shown in Fig.\ref{fig1}. The structural randomness does not affect the arrangement and orientation of the magnetic spins, therefore the metallic glassy magnets is possible. In a rhombus unit we consider the interactions $J$ between two neighbor spins and $J_{\text{P}}$ between the diagonal spin pairs. The Hamiltonian of the system is given by 
\begin{equation}
E=-J\underset{<i,j>}{\sum}S_{i}S_{j}-H_{i}{\textstyle\sum}S_{i}-J_{\text{P}}\underset{<j,j'>}{\sum}(\mathcal{A}(\theta_{i})\cdot S_{j}S_{j'}), \label{Hamiltonian}
\end{equation}
where $<S_{i},S_{j}>$ is the neighbor pair, $<S_{j},S_{j'}>$ is the diagonal pair, $H$ is magnetic field, $\theta_{i}$ is the angle against the diagonal pair, and $\mathcal{A}$ is the angle energy coefficient. 

Assume that $J$ is fixed and the angle only affects the diagonal interaction, then by adjusting $J_{\text{P}}$ we can have a system with particular randomness away from the ideal crystal state.
And further assume that the deviation only affect the diagonal interaction in the rhombus due to the diagonal length change, and it is linearly proportional to the diagonal length, the we have the diagonal interaction energy:
\begin{equation}
E_{p}(\theta)=-J_{p}\cdot S_{j}S_{j'}\cdot\frac{\sqrt{2(1-\cos\theta)}}{\sqrt{2(1-\cos\theta_{o})}},
\end{equation}
where $\theta_{o}$ is the ground state angle with the lowest energy, it is set to be $\pi/2$ in this paper. Then we have $\mathcal{A}(\theta)$ as the angle energy coefficient:
\begin{equation}
\mathcal{A}(\theta)=\sqrt{\frac{1-\cos\theta}{1-\cos\theta_{o}}}, \label{eq:A}
\end{equation}

Imagine the entire lattice has an original site $S_{0}$ on the $0$th level, then the sites nearest to $S_{0}$ are marked as $S_{1}$ (or ${S'}_{1}$ to distinguish the two sites in the same unit), the site diagonal to $S_{0}$ is marked as $S_{2}$, and so on. On one rhombus unit there are three sub-trees $\mathcal{T}_{n+1}$ and $\mathcal{T}_{n+2}$ contributing to the sites $S_{n}$, ${S'}_{n+1}$ and $S_{n+2}$, and a larger sub-tree $\mathcal{T}_{n}$ is synthesized by adding the weights of the local unit. With the particular spin state of $S_{n}=\pm1$, define the partial partition function (PPF) $Z_{n}(S_{n})$ on level $n$ to count the contribution of sub-tree to the site $S_{n}$ (not included), then the $n$th level PPF is the function of PPFs on the previous level $n+1$, $n+2$ and the weights of local unit $w(\gamma)$:

\begin{equation}
Z_{n}(+)=\underset{\gamma=1}{\overset{8}{\sum}}Z_{n+1}(S_{n+1})Z_{n+1}(S_{n+1}^{\prime})Z_{n+2}%
(S_{n+2})w(\gamma),\label{eq:PPF+_Recursion}
\end{equation}
\begin{equation}
Z_{n}(-)=\underset{\gamma=9}{\overset{16}{\sum}}Z_{n+1}(S_{n+1})Z_{n+1}(S_{n+1}^{\prime})Z_{n+2}(S_{n+2})w(\gamma). \label{eq:PPF-_Recursion}
\end{equation}   
And the partition function (PF) of the whole sub-tree on $n$th level is given by
\begin{equation}
Z_{n}=Z_{n}(+)e^{\beta HS_{n}}+Z_{n}(-)e^{-\beta HS_{n}},\label{eq:PF}
\end{equation}
which summers all the weights at level $n$ including the spin $S_{n}$.

Then this model can be exactly solved by the recursive relations between the PF and PPFs on successive levels to obtain the probability of spins' state at particular temperature, and the magnetization and thermodynamics can be derived from the probability. For the calculation details please refer to previous works on the Ising model on Husimi lattice \cite{PDG5,arxiv1,ran1,jpsj} and the supplemental materials.

\section{Results and Discussion}
In experimental the magnetization of BMG is usually induced by an external magnetic field, nevertheless in the 2D Ising model \cite{Onsager}, spins can spontaneously transit to uniform orientation without external field. And the temperature where this spontaneous magnetization occurs can be taken as the Curie temperature \cite{SLAWSKA-WANIEWSKA}. In this section, all the discussions on the magnetization and thermodynamics refer to the case of spontaneous magnetization.
\subsection{The magnetization and $T_{C}$}

\begin{figure}
    \centering{
	\includegraphics[width=0.48\textwidth]{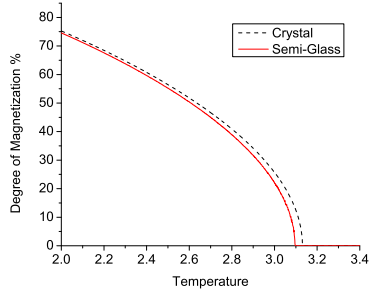}
	\caption{The degree of magnetization $M$ of Ising spins on random-angled Husimi lattice comparing to the regular Ising magnets on fixed lattice, with $J=1$ and $J_{\text{P}}=0.2$. The temperature is normalized with Boltzmann constant $k_{\text{B}}=1$.}
	\label{fig2}
    }
\end{figure}

While the magnetization is usually characterized by the magnetic field strength in the real materials, in the theoretical model of spins ensemble the magnetization can be simply presented by the probability of spins orientation. This probability of Ising spins can be solved as the `solution' of the model. The solutions on fixed Husimi lattice (crystal) and random lattice (semi-glass) with $J=1$ and $J_{\text{P}}=0.2$ are shown in the Fig. \ref{fig2}. Note that the temperature in this work is normalized with Boltzmann constant $k_{\text{B}}=1$. Several observations can be drawn here: 1) At high temperature, one spin has fifty percent probability to be in either $+1$ or $-1$ state, then the average orientation is 0; 2) With the cooling process, there occurs the spontaneous magnetization in both systems, and the Curie temperature is $T_{C}=3.131$ and $3.097$ for crystal and semi-glass; 3) Below $T_{C}$, the magnetization continue to increase with the decrease of temperature, and at zero temperature, all the spins will be in the same state and we then have a 100\% magnetization on the lattice (not shown in the Fig.\ref{fig2}). It is clear that the semi-glass system is less stable and has a lower $T_{C}$, i.e. the system is easier to proceed a transition, and the magnetization of semi-glass is weaker at a particular temperature comparing to the crystal.

On the other hand, the $T_{C}$ reduction is relatively small comparing the experimental results. This is because that our model represents a semi-glass as mentioned above, and the parameter $J_{\text{P}}=0.2$ in Fig.\ref{fig2} is merely a reference setup. We can arbitrarily obtain a particular reduction to fit the real system by adjusting the parameter, the details will be discussed below. 

\subsection{Paramagnetic and ferromagnetic behavior with external field $ H $}

In experimental, the magnetization responding to the applied external field is a critical characterization, e.g. the VSM test, to reveal the magnetic properties. In our theoretical model, the field $H$ in Eq.\ref{Hamiltonian} can serve as an adjustment option for the same purpose. Figure \ref{fig2}a presents the 3D mapping of the magnetization with the variation of external field $H$ and temperature $T$. The paramagnetic and ferromagnetic ranges differentiated by the critical temperature can be observed, the sign of magnetization indicates the orientation of spins magnetism, and the bold discontinuous line is the spontaneous magnetization with $H=0$ shown in Fig.\ref{fig2}. 

\begin{figure}
    \centering{
    \subfloat[]{
    \includegraphics[width=0.48\textwidth]{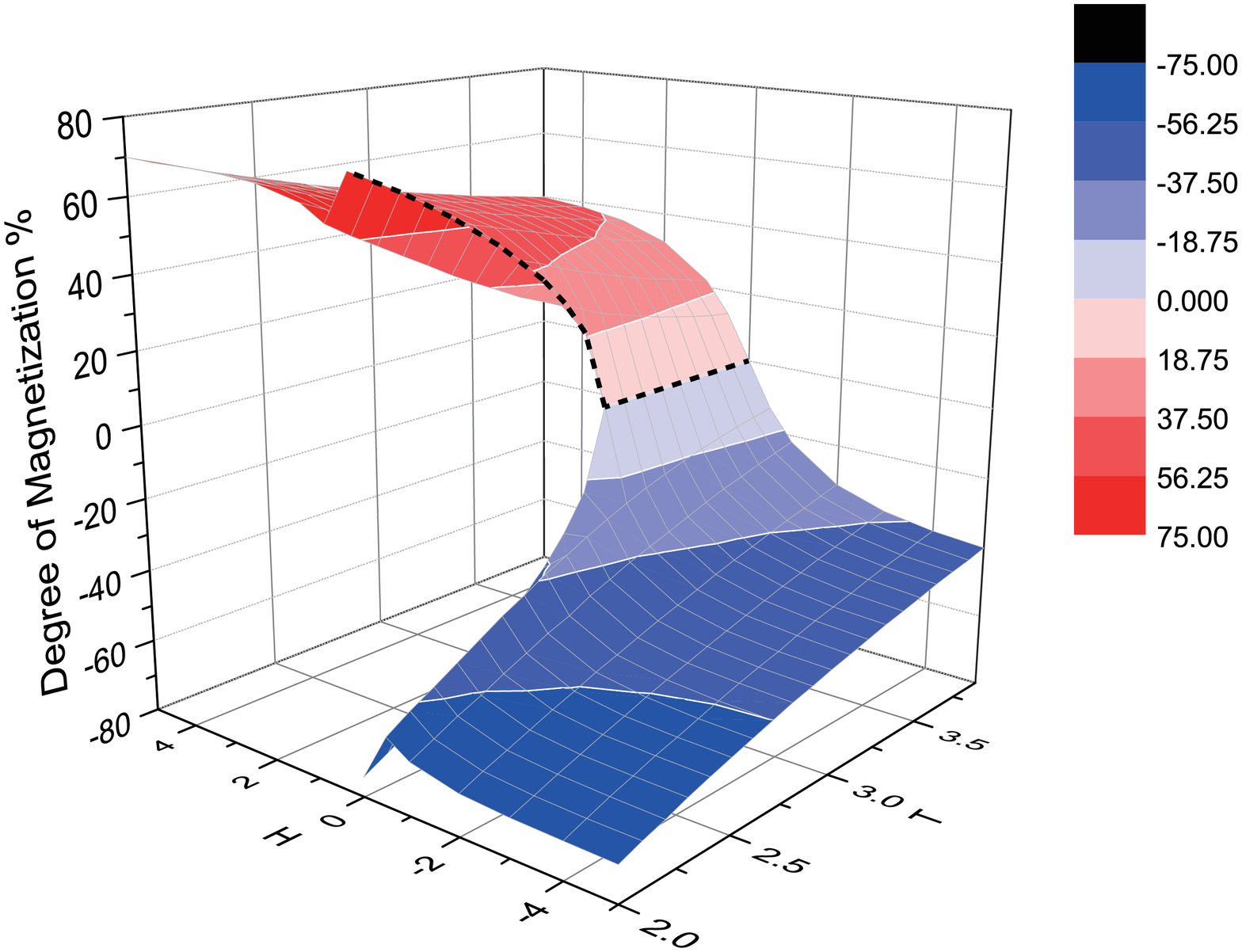}
    }
    \subfloat[]{
	\includegraphics[width=0.48\textwidth]{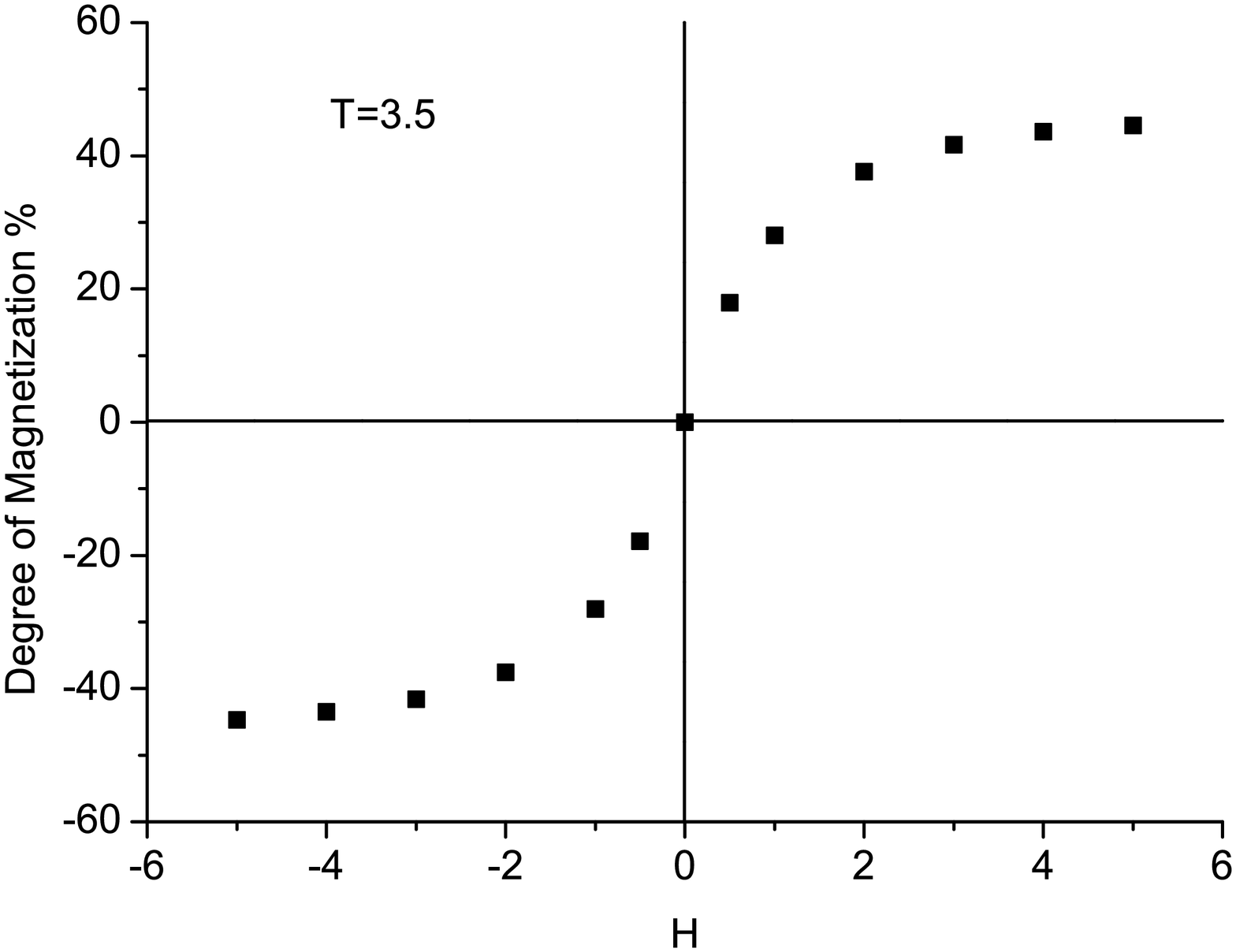}
	}\\
	\subfloat[]{
	\includegraphics[width=0.48\textwidth]{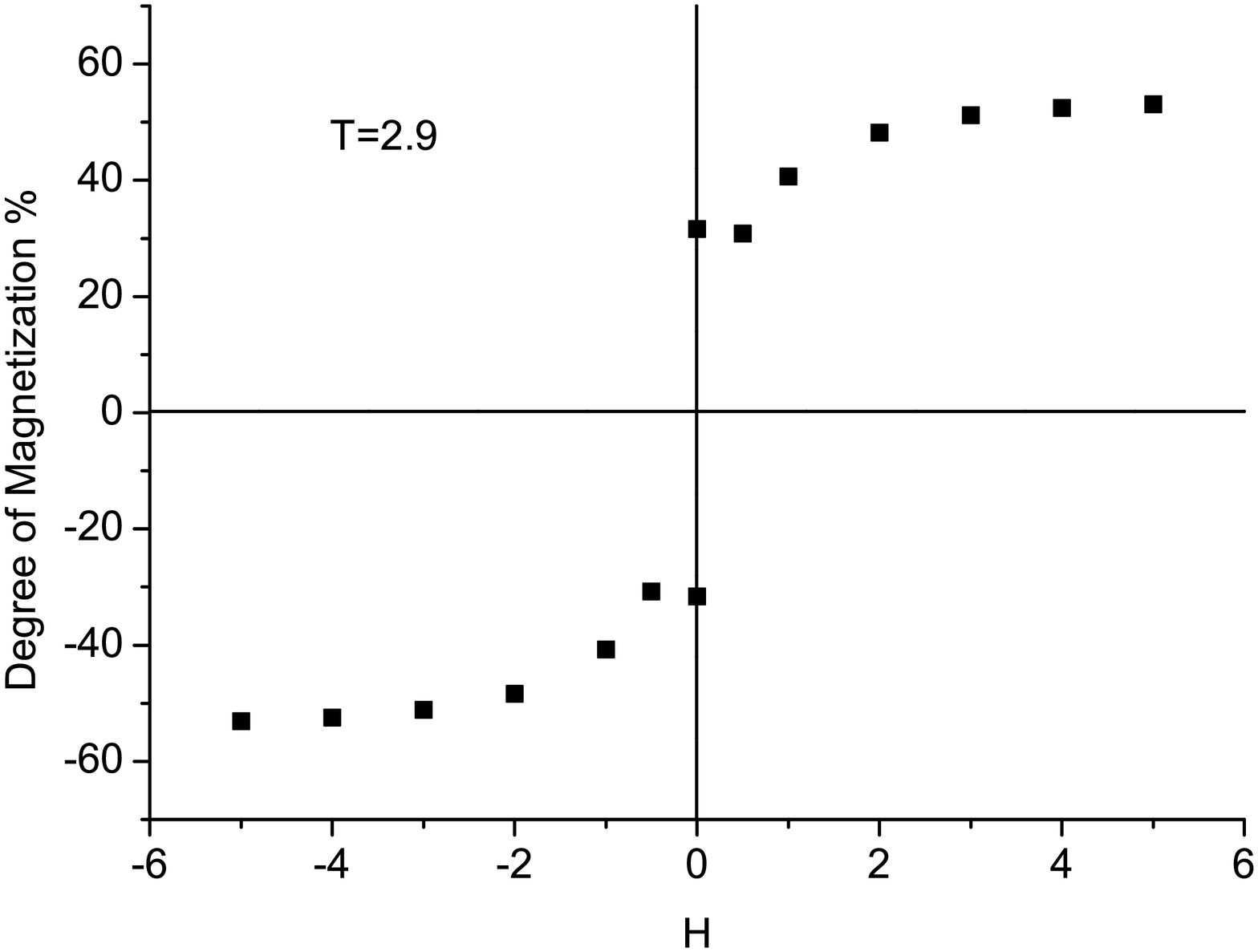}
	}
	\subfloat[]{
	\includegraphics[width=0.48\textwidth]{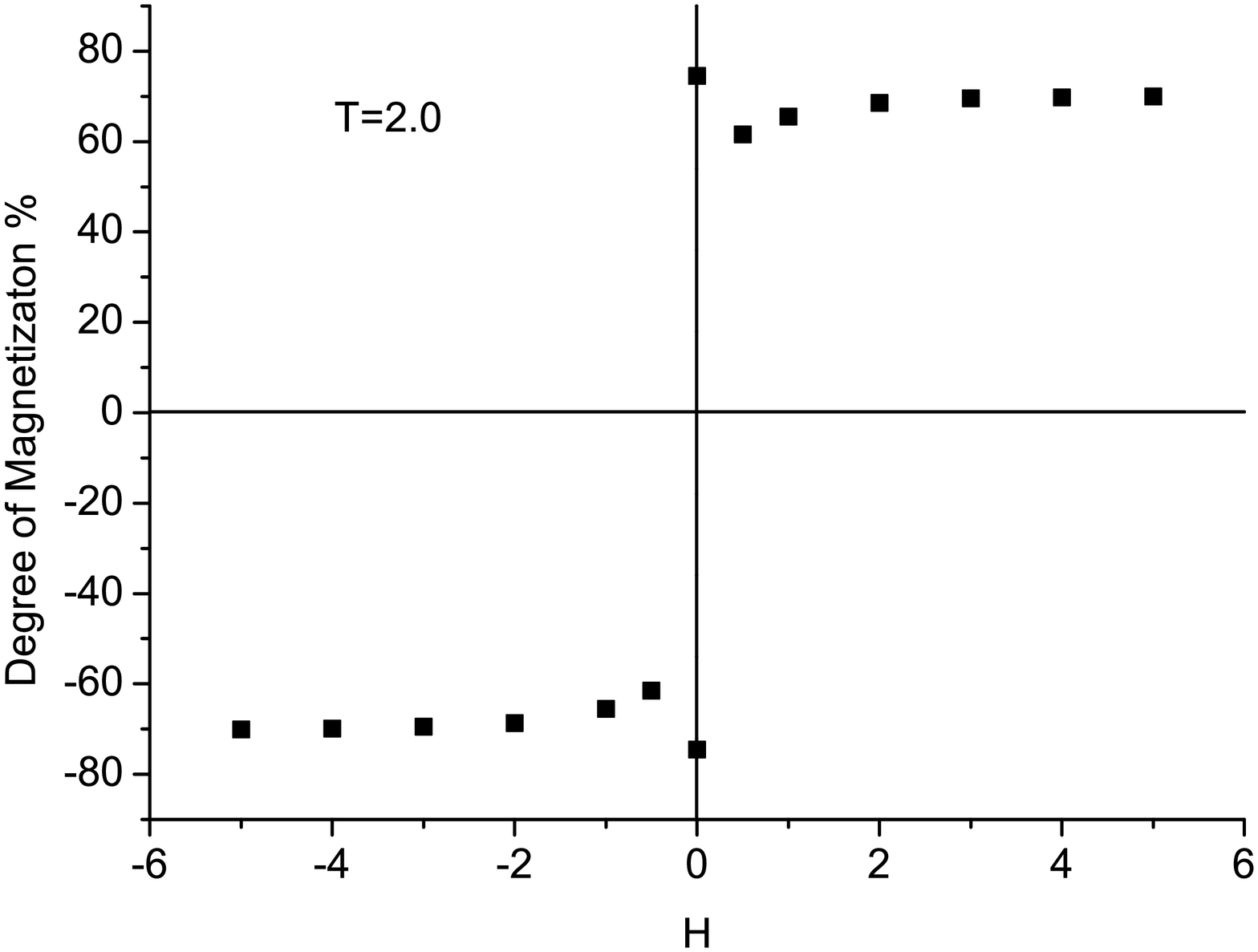}
	}
	\caption{(a) The 3D mapping of magnetizations with various external field $H$ and temperature $T$. (b, c, d) The magnetization curve at $T=3.5$, $2.9$ and $2.0$. }
	\label{fig3}
    }
\end{figure}

Three magnetization curves at $T=3.5$, $2.9$ and $2.0$ are extracted out and presented in Fig.\ref{fig2}b, c and d. While in Fig.\ref{fig2}b the paramagnetic behavior is clear at high $T$ as expected, the magnetic behavior below $T_C$ is somehow bizarre, that the spontaneous magnetization is higher than the system with small field applied ($H=0.5$) in Fig.\ref{fig3}c, and at lower temperature this phenomenon is more distinct as shown in Fig.\ref{fig3}d. The exact reason of this phenomenon is hypothesized as a defective feature of our semi-glass model: Since the model is static, the hysteresis loop cannot be simulated, and for the ferromagnetic case we can only infer the coercivity by the free energy calculation of self-magnetization case with $H=0$. While in Fig.\ref{fig3}c and d the extrapolations with $H\rightarrow0$ have a clear trend pointing to the origin, implying a soft magnetic behavior even below $T_C$, which is how a real magnetic metallic glass exhibits. Nevertheless, nor like the real system where the conformational randomness can cancel out the orientation of magnetic spins, the off-crystal structure of the 2D lattice is fundamentally independent of the spins orientation, subsequently the amorphousness presented by random conformation cannot cancel the magnetization. In another word, regardless of how the rhombus unit is twisted with higher angling energy, the strong coupling between spins at low $T$ is not affected. Therefore, at low temperature the intrinsic magnetism from spontaneously oriented spins will resist the soft magnetic behavior and retains a firm coercivity, while the application of external field (even a small value) will dominate the spins orientation and simulate the real behavior of metallic glass in a field.  

\begin{figure}
    \centering{
	\includegraphics[width=0.48\textwidth]{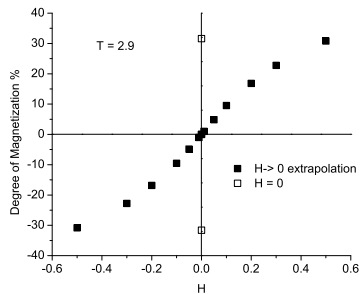}
	\caption{The variation of magnetization $M$ with the $H\rightarrow0$ in the near-zero range in the case of $T=2.9$ (Fig.3c). The system shows exact soft magnetism with the presence of $H$, while the self-magnetization case gives a meta-stable solutions with $H=0$.}
	\label{fig4}
    }
\end{figure}

To illustrate the soft magnetism with the presence of the external magnetic field, a detailed $M$ vs. $H$ behavior of the near-zero range in the case of Fig.\ref{fig3}c is presented in the Fig.\ref{fig4}. It is clear to observe that the $H\rightarrow0$ corresponds to $M\rightarrow0$, as an extreme value of $H=1\times10^{-6}$ raises $M=9.84\times10^{-5}$. However the calculation of pure self-magnetization case distinguishably gives the solutions out of the extrapolation, which is the $M=\pm 34.2\%$ with $H=0$ in Fig.\ref{fig4}. Notwithstanding, besides this defect, the application of external field on this model to simulate the magnetization test is still practical to describe the magnetic metallic glass.

\subsection{The thermodynamics}

While various electromagnetic and thermodynamic properties of the glassy ferromagnet are being actively researched, the study and report on entropy is rare. A probable reason of this lack of attention maybe due to that the entropy is not a directly measurable variable. Nevertheless, considering that this research field is fundamentally based on the natural difference between crystal and amorphous state, should entropy be an important quantity to pay attention on. Some experimental work reported the positive correlation between higher entropy with enhanced GFA or magnetization \cite{entropy}. 

\begin{figure}
    \centering{
    \subfloat[]{
	\includegraphics[width=0.48\textwidth]{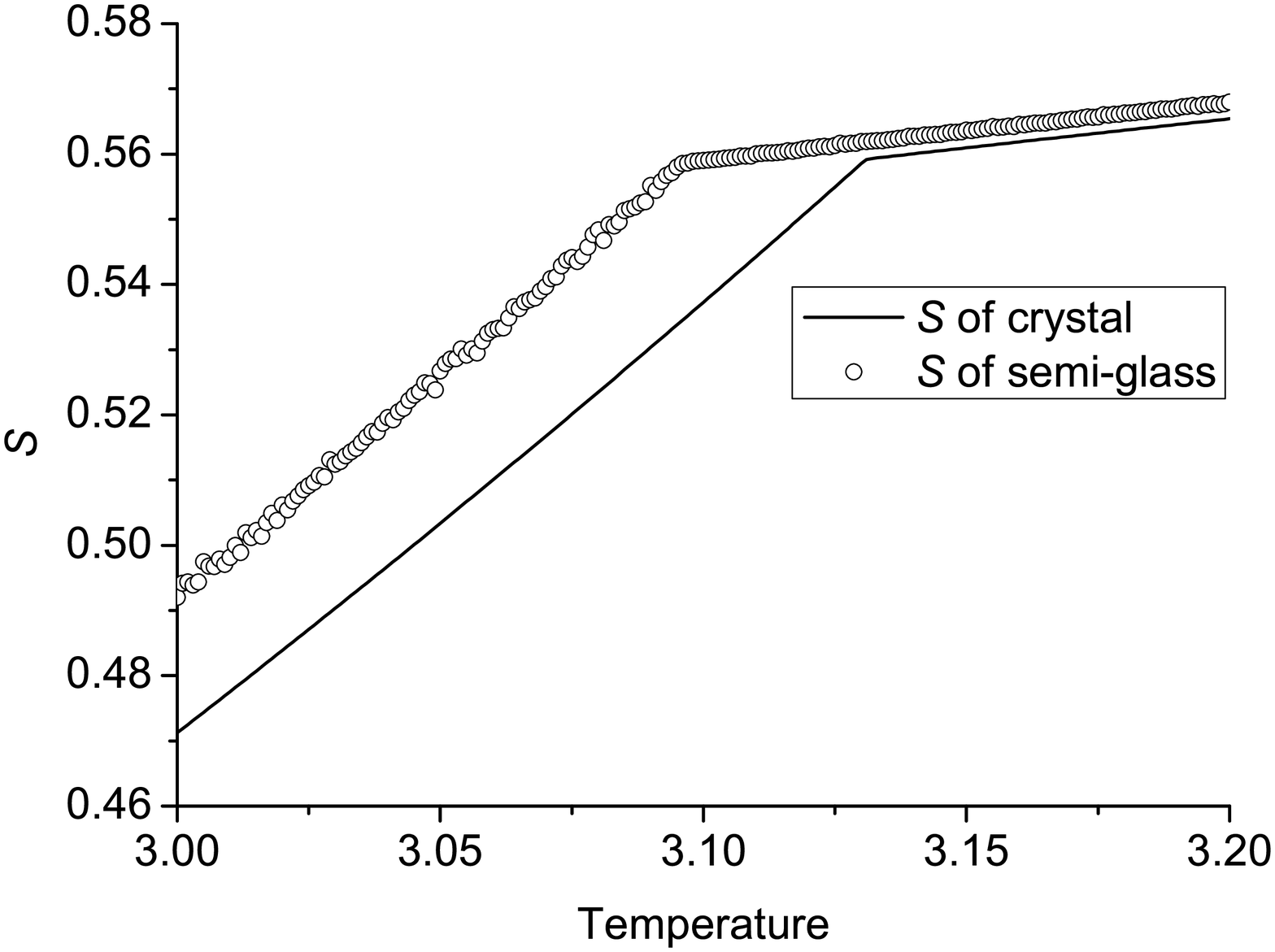}
	}\\
	\subfloat[]{
	\includegraphics[width=0.48\textwidth]{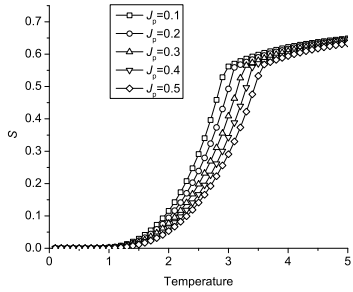}
	}
	\caption{(a) The entropy of Ising magnet on the random-angled Husimi lattice (semi-glass) comparing to the regular Ising magnet (crystal). (b) Entropy variation with diagonal interaction parameter $J_{\text{P}}$.}
	\label{fig5}
    }
\end{figure}

Unlike experimental, in the theoretical calculation the entropy is easy to be monitored. Figure \ref{fig5}a shows the comparison between the entropy of Ising magnet on random rhombus lattice and in the crystal form around the $T_{C}$ region. For both cases, the transitions on entropy at $T_{C}$ are clear, and accords well to their spontaneous magnetization revealed in Fig.\ref{fig2}.

In the Hamiltonian (Eq.\ref{Hamiltonian}), the diagonal interaction parameter $J_{\text{P}}$ is the only variable related to the randomness of angle, and $J_{\text{P}}$ also has positive value setup in the model like $J$, that is, higher $J_{\text{P}}$ prefers same spins orientation and cancels the effect of randomness, therefore we can take $J_{\text{P}}$ to be the inverse randomness parameter, i.e. the lower the $J_{\text{P}}$ is, the system is more amorphous and farther away from the crystal. The effects of $J_{\text{P}}$ variation on the $T_{C}$ and entropy are shown in Fig. \ref{fig5}b. While all the systems exhibit similar behaviors, the variation of $J_{\text{P}}$ by 0.1 from 0.1 to 0.5 give the $T_{C}$ as 2.990, 3.097, 3.296, 3.401, and 3.598 respectively, exhibiting a rough linear relationship. It is easy to observe that at a particular temperature, the increase of $J_{\text{P}}$ corresponds to a more stable system with lower GFA, smaller entropy and higher magnetization. This theoretical finding agrees well with others' experimental observations.

\section{Conclusion}

The Ising spins were modeled on a Husimi rhombus lattice with random angles to describe the ferromagnetic metallic glass. A diagonal interaction parameter $J_{\text{P}}$ related to the angle randomness was added into the Hamiltonian to disturb the fixed configuration, subsequently provided a semi-glass deviated from ideal crystal. The model was exactly solved and exhibits the spontaneous magnetization with transition temperature $T_{C}$. The glassy system presents a lower $T_{C}$ and weaker magnetization than the ferromagnetic crystal. Both paramagnetic and ferromagnetic behaviors at various $T$ with external field to mimic the magnetization test were investigated, with a general expectable observations, that the system behaves as ideal soft-magnetism with the presence of external field $H$, however a higher spontaneous magnetization was also found as a meta-stable solution of the pure self-magnetization system with $H=0$. The entropy and the effect of parameter $J_{\text{P}}$ were studied, the observation agrees well with others' experimental findings, that the thermodynamically less stable system corresponds to a higher GFA, larger entropy and weaker magnetization.

\end{document}